\documentclass[aip, reprint, dcolumn, longbibliography]{revtex4-1}
\usepackage[]{graphicx}
\usepackage{caption}
\usepackage{floatrow}
\usepackage{amssymb}
\usepackage{amsmath}
\usepackage{soul}
\usepackage{xcolor}
\usepackage{lipsum}
\usepackage{empheq}
\DeclareMathOperator*{\argmin}{argmin}

\begin{document}

\title{Determining the refractive index, absolute thickness and local slope of a thin transparent film using multi-wavelength and multi-incident-angle interference}

\author{Mengfei He}
\author{Sidney R. Nagel}
\affiliation{Department of Physics, The James Franck and Enrico Fermi Institutes, The University of Chicago, Chicago, Illinois, 60637}

\begin{abstract}  
	We describe a high-speed interferometric method, using multiple angles of incidence and multiple wavelengths, to measure the absolute thickness, tilt, the local angle between the surfaces, and the refractive index of a fluctuating transparent wedge.  The method is well suited for biological, fluid and industrial applications.
\end{abstract}
\maketitle

\section{Introduction}
It is well known that the fringes caused by the interference of light reflecting from the front and back surfaces of a transparent film can be used to determine the \textit{relative} thickness of the film as a function of position.  Interference fringes can also be used to determine the film's \textit{absolute} thickness.  The optical path difference between the two reflected rays, $\Delta L$, is determined by the film thickness, $h$, and the angle of incidence, $\theta$.  The absolute thickness $h$ can be determined by measuring the variation of 
$\Delta L$ as a function of $\theta$.

The general principle of probing the structural information perpendicular to a surface through interference with a varying incident angle dates back to Kiessig~\cite{kiessig1931}.  It has been applied in X-ray and neutron reflectometry to characterize solid structures such as the electron distribution and oxidation depth~\cite{zhou1995, parratt1954, russell1990, stoev1997,majkrzak1991, ankner1999,dura1998}.  The setup usually contains a mechanism that moves in order to change the angle of incidence~\cite{takabayashi1987, ishikawa2004,choi2010}.  The necessity of mechanically changing the incident angle limits such measurements to static structures that do not vary rapidly over time.  Moreover, the design and accompanying derivations for such techniques are usually limited to structures with parallel layers.

In this paper, we present a method to measure the thickness of a transparent layer whose thickness \textit{fluctuates} in both space and time.  We adapt a technique using an impinging spherical wave 
first introduced by Gold \textit{et al}~\cite{gold1991,gold1991_2} to achieve a range of angles of incidence simultaneously.  Combined with high-speed imaging to compare frames, we demonstrate that this technique allows the \textit{instantaneous} absolute thickness as well as the refractive index of the film to be determined from the interference-fringe geometry; using multiple wavelengths simultaneously improves precision.  More importantly, we show that this technique can be used to measure not only the film thickness at a point but also  the thickness gradient of a film that does not have parallel surfaces as well as its overall tilt.   This approach extends traditional solid thin-film measurements to dynamic regimes such as fluid layers and fluctuating biological membranes. 

\section{Interference at the focal plane}

Because the geometry of this experiment is a bit complicated, we will describe the technique in four stages. (i)  We will first show how the technique of Gold \textit{et al}~\cite{gold1991,gold1991_2} can be adapted to rely only on measuring the geometry of the interference pattern (without relying on measuring the intensity).  We determine the index of refraction as well as the thickness of a film with parallel surfaces (so that the opening angle, $\alpha$, is zero) and which is oriented perpendicularly to the observation axis (so that the tilt angle $\gamma = 0$). (ii)  We will next analyze the case where the sample with parallel surfaces ($\alpha = 0$) is tilted from the previous case by $\gamma \ne 0$. (iii)  We then treat the case where there is no tilt ($\gamma = 0$), but the surfaces of the film are not parallel and have a local opening angle $\alpha \ne0$.  (iv)  Finally, we will show how these results can be used to determine the local thickness gradient of the wedge as well as an overall tilt of the sample in an arbitrary direction.  

\subsection{Sample with parallel surfaces and no tilt ($\alpha=\gamma = 0$)}

\begin{figure}[H]
	\includegraphics[width=0.9\textwidth]{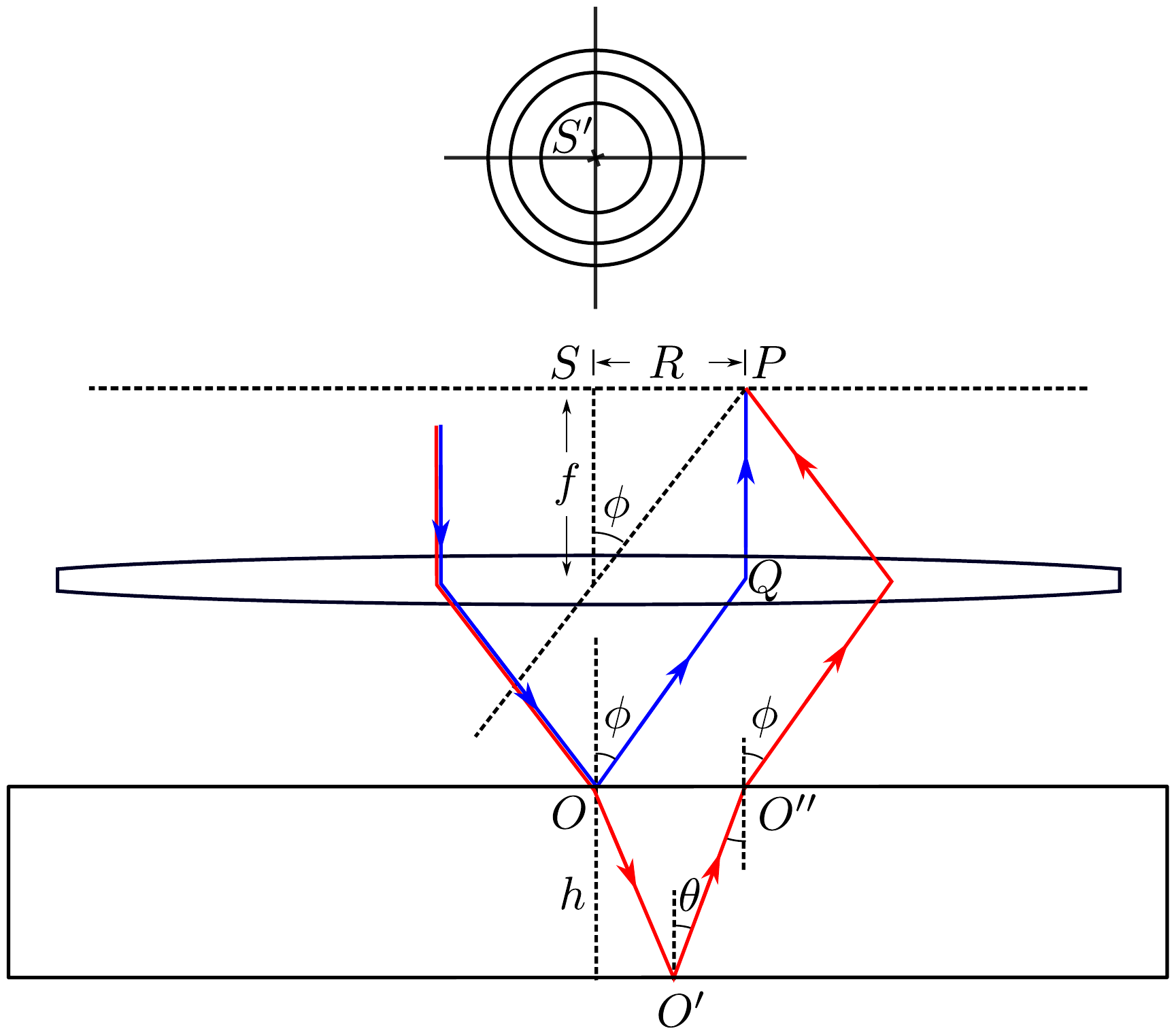}
	\caption{Setup for measuring thickness and refractive index of a parallel slab with no tilt ($\alpha=\gamma=0$).  Rays marked by blue and red meet at the upper focal plane (dotted line) of the lens.  This leads to the interference pattern of concentric circles in the center $S$ of the focal plane (as shown schematically above that plane).}
	\label{fig:haidinger_a0g0}
\end{figure}

The simplest situation consists of a parallel slab ($\alpha=0$) with no tilt ($\gamma = 0$).  As considered by Gold \textit{et al}~\cite{gold1991,gold1991_2} and recently by Kim \textit{et al}~\cite{kim2017} and Kim \textit{et al}~\cite{kim2018}, the sample is placed at a focal length, $f$, below a convex lens (Fig.~\ref{fig:haidinger_a0g0}).  A parallel beam of light of wavelength, $\lambda$, is focused by the lens onto the upper surface of the film.  The light will be reflected by both the upper and lower surfaces at points $O$ and $O'$ as shown.  The reflected light will return through the lens, and converge at point $P$ on the focal plane of the lens in the region on the side away from the film. The angle $\phi$ between the central axis and the line from the lens center to point $P$ is  
\begin{align}
	\phi = \arctan \frac{R}{f},
	\label{eq:geometry}
\end{align}
where $R$ is the distance from $P$ to $S$.

It is easy to show that the two different paths, marked by blue and red in Fig.~\ref{fig:haidinger_a0g0}, have an optical path difference
\begin{align}
	\label{eq:opd}
	\Delta L = 2nh\cos\theta,
\end{align}
where $h$ is the wedge thickness at the point $O$, $\theta$ is the angle of incidence at the bottom interface at $O'$ and $n$ is the refractive index of the film.  Including the $\pi$ phase change at reflections (assuming near normal incidence), there will be destructive/constructive interference, corresponding to dark/bright fringes if:
\begin{subequations}\label{eq:deconstructive}
\begin{align}
	\label{eq:destructive}
	\text{destructive:} \hspace{1em} &2nh\cos\theta_{m} = m\lambda, \\
	\label{eq:constructive}
	\text{constructive:} \hspace{1em} &2nh\cos\theta_{m+\frac{1}{2}} = (m+\frac{1}{2})\lambda, 
\end{align}
\end{subequations}
where $m$ is an integer indicating the order of interference. 

Taking into account the three dimensions of the setup, there will be bright and dark circular interference patterns at the top focal plane.  (This pattern is referred to as ``Haidinger's fringes''~\cite{rayleigh1906, raman1940} to be distinguished from the more familiar ``Newton's fringes''. Haidinger's fringes occur because of a variation of the angle of incidence of the light source whereas Newton's rings result from a variation of the sample thickness.)  Due to symmetry (\textit{i.e.}, $\alpha = \gamma = 0$), the interference pattern at the focal plane consists of circular, concentric rings. The pattern center $S'$ coincides with the screen center $S$, as is schematically shown at the top of Fig.~\ref{fig:haidinger_a0g0}.  

Using Snell's law at $O''$,
\begin{align}
	\label{eq:snell_a0g0}
	\sin\phi = n\sin\theta.
\end{align}
Eqs.~\ref{eq:deconstructive} reduce to: 
\begin{subequations}\label{eq:deconstructive_a0g0}
\begin{align}
	\label{eq:destructive_a0g0_1}
	\text{destructive:} \hspace{1em} &2nh\sqrt{1-\frac{\sin^2{\phi_{m}}}{n^2}} = m\lambda,\\
	\label{eq:constructive_a0g0_1}
	\text{constructive:} \hspace{1em} &2nh\sqrt{1-\frac{\sin^2{\phi_{m+\frac{1}{2}}}}{n^2}} = (m+\frac{1}{2})\lambda.
\end{align}
\end{subequations}

For the fringes closest to the center, $\sin^2\phi/n^2 \ll 1$.  (For $n\sim1.2$, a distance $R=10$ mm and a focal length $f=50$ mm, the fringes correspond to $\sin^2\phi/n^2 \sim 0.03$.)  
Thus for the inner fringes Eqs.~\ref{eq:destructive_a0g0_1} and~\ref{eq:constructive_a0g0_1} can be approximated as:
\begin{subequations}
	\label{eq:deconstructive_a0g0_2}
\begin{align}
	\label{eq:destructive_a0g0_2}
	\text{destructive:} \hspace{1em} &2nh(1-\frac{\sin^2{\phi_{m}}}{2n^2}) = m\lambda,\\
	\label{eq:constructive_a0g0_2}
	\text{constructive:} \hspace{1em} &2nh(1-\frac{\sin^2{\phi_{m+\frac{1}{2}}}}{2n^2}) = (m+\frac{1}{2})\lambda.
\end{align}
\end{subequations}

The difficulty of accurately identifying $m$~\cite{choi2010,park2017,kim2017} can be avoided by cancelling $m$ from Eqs.~\ref{eq:destructive_a0g0_2} and~\ref{eq:constructive_a0g0_2}:
\begin{align}
	\label{eq:hn}
	\frac{1}{\lambda}[\cos(2\phi_{m+\frac{1}{2}})&-\cos(2\phi_{m})] =\frac{n}{h}.
\end{align}
Thus, measuring the fringe positions, $\phi_m$'s, and calculating the mean quantities of $[\cos(2\phi_{m+1/2})-\cos(2\phi_{m})]/\lambda$ determines $n/h$.

\subsubsection{Measuring the thickness/refractive index}
In special cases where $n$ is known, such as in an air gap ($n=1$) or a water film ($n=1.33$), this method alone provides a precise way of measuring the absolute thickness of $h$.  Likewise, if the sample fluid can be sandwiched in a gap of a known thickness (ensuring $\alpha=\gamma=0$), the refractive index can be precisely measured.

To check this experimentally, we measured the thicknesses of 3 different pieces of glass (denoted as $h\textsubscript{g0}$, $h\textsubscript{g1}$, $h\textsubscript{g2}$) with known refractive indices, and the refractive indices of air ($n\textsubscript{a}$), water ($n\textsubscript{w}$), and a water/glyceral mixture ($n\textsubscript{wg}$, glycerol mass fraction 53.5 \%) using cells with known gap thicknesses.  Table~\ref{tab:h_and_n} shows the collected results.  The ``Micrometer" values of thicknesses were measured by either a MITUTOYO micrometer or a MITUTOYO caliper.  The ``Literature" values of refractive indices were taken from literature~\cite{hoyt1934, born2013principles}.  The errors quoted represent the typical differences between two separate interference measurements. 

\begin{table}[H]
\begin{tabular}{ | c | c | c | } 
\hline
& Micrometer/Literature & Interference measurement\\ 
\hline
$h\textsubscript{g0}$ ($\mu$m)& 203  & 205 $\pm$ 0.5\\ 
\hline
$h\textsubscript{g1}$ ($\mu$m)& 1050  & 1030 $\pm$ 9\\ 
\hline
$h\textsubscript{g2}$ ($\mu$m)& 6071 & 6061 $\pm$ 17\\ 
\hline
$n\textsubscript{a}$ & 1.0003~\cite{born2013principles} & 1.017 $\pm$ 0.004\\ 
\hline
$n\textsubscript{w}$ & 1.33303~\cite{hoyt1934, born2013principles} & 1.3362 $\pm$ 0.0004\\ 
\hline
$n\textsubscript{wg}$ & 1.404~\cite{hoyt1934} & 1.39 $\pm$ 0.01\\ 
\hline
\end{tabular}
\caption{Measured thicknesses for different pieces of glass ($h\textsubscript{g0}$, $h\textsubscript{g1}$, $h\textsubscript{g2}$), and refractive indices for air ($n\textsubscript{a}$), water ($n\textsubscript{w}$), and a water/glyceral mixture ($n\textsubscript{wg}$, glycerol mass fraction 53.5 \%).  ``Micrometer'' values of $h$'s were  measured mechanically with a micrometer or caliper.  ``Literature'' values of $n$'s were taken from literature~\cite{hoyt1934, born2013principles}.  The values obtained by the interference measurement are shown in the right column.}
\label{tab:h_and_n}
\end{table}

\subsubsection{Measuring the refractive index in a fluctuating film}

In the more general situation where neither $h$ nor $n$ are known beforehand, they can be separated if $h$ varies with time $t$: $h=h(t)$.  This means the interference pattern will expand/shrink over time, which can be tracked using high-speed video.  Using Eq.~\ref{eq:hn}, the quotient, $n/h(t)$, can be obtained continuously as a function of $t$.  On the other hand, the variation of the product, $nh(t)$, can simultaneously be found as the order of the interference in the pattern center changes by $\Delta p$ fringes.  This is because from Eq.~\ref{eq:opd}, the center of the pattern corresponds to the maximum optical path difference at $\theta=0$:
\begin{align}
	\Delta L_{\text{max}} = 2nh.
\end{align}
Thus we have  
\begin{align}
	\label{eq:ht}
	2nh(t+\Delta t) - 2nh(t) = \Delta p\lambda,
\end{align}
where $\Delta p$ is the number of fringes sinking into (negative) or emerging from (positive) the center of the pattern during the time interval $\Delta t$.  Therefore, we have:

\begin{align}
	\label{eq:hnt}
	\frac{h(t+\Delta t)}{n} - \frac{h(t)}{n} = \frac{\Delta p\lambda}{2n^2},
\end{align}
or,
\begin{align}
	\label{eq:hp}
	\frac{\Delta \frac{h}{n}}{\Delta p} = \frac{\lambda}{2n^2}.
\end{align}
Therefore, a linear regression between $h(t)/n$ and the \emph{accumulated} number of fringes, $p$,  that have appeared/disappeared for an extended period of time results in a slope of $\lambda/2n^{2}$, from which $n$ can be accurately deduced.   

\begin{figure}[H]
	\includegraphics[width = 0.9\textwidth]{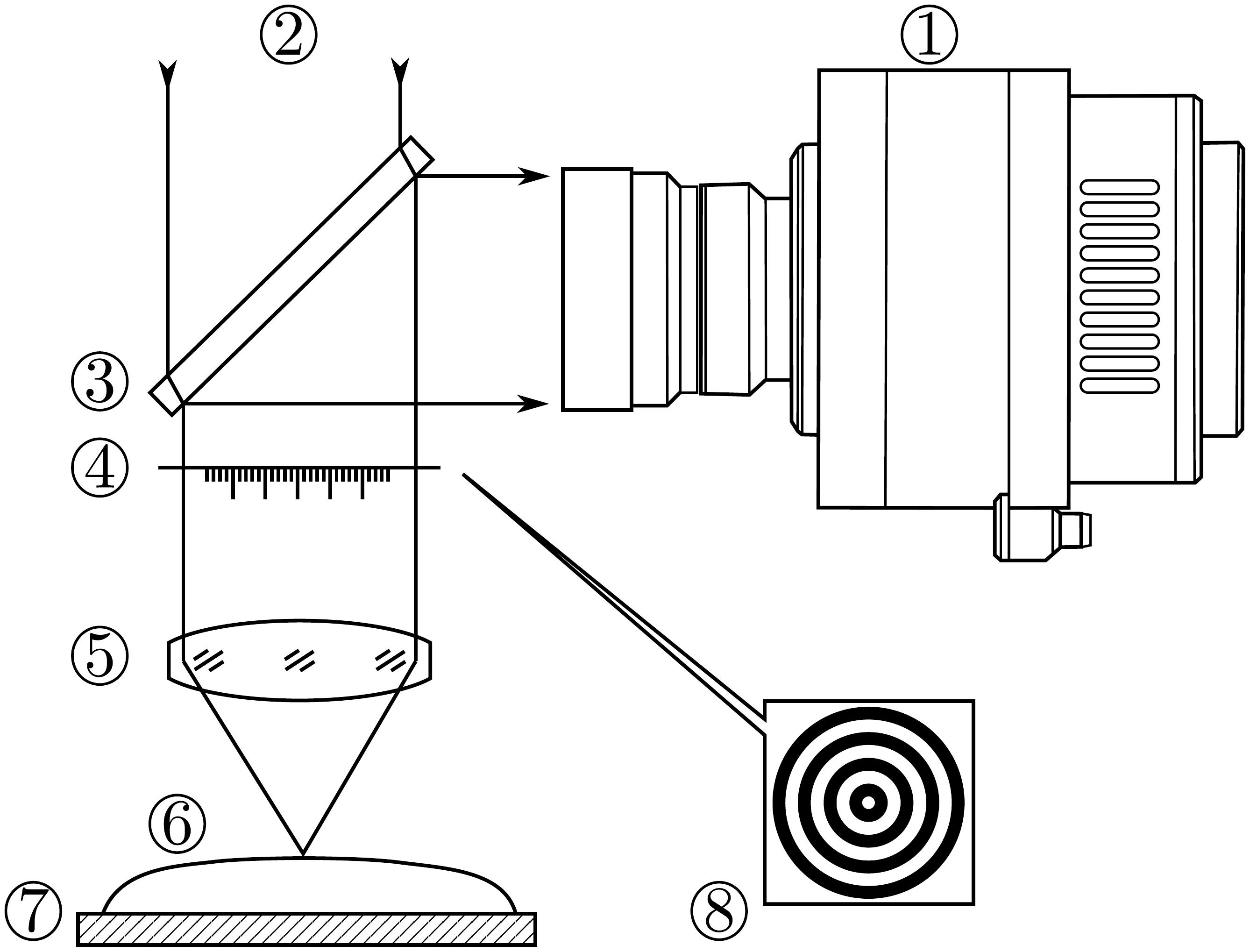}
	\caption{Experimental setup for measuring a fluctuating liquid layer.  \textcircled{\raisebox{-.9pt} {1}}: high-speed camera;  \textcircled{\raisebox{-.9pt} {2}}: laser beam ($\lambda=$ 532 nm);  \textcircled{\raisebox{-.9pt} {3}}: beam splitter;  \textcircled{\raisebox{-.9pt} {4}}: focal plane + reference scale;  \textcircled{\raisebox{-.9pt} {5}}: convex lens (Nikon f=50mm, 1:1.8);  \textcircled{\raisebox{-.9pt} {6}}: liquid layer;  \textcircled{\raisebox{-.9pt} {7}}: substrate;  \textcircled{\raisebox{-.9pt} {8}}: interference pattern.}
	\label{fig:setup}
\end{figure}

We have applied this method to measure the refractive index of a liquid film, using the setup shown schematically in Fig.~\ref{fig:setup}.  The substrate (\textcircled{\raisebox{-.9pt} {7}}) was made to move so that the liquid film was fluctuating in its thickness.  We used a Nikon lens (f=50mm, 1:1.8) in the place of the convex lens (\textcircled{\raisebox{-.9pt} {5}}) to make a laser beam of wavelength $\lambda=532$ nm converge on the upper interface of the liquid film.  The interference pattern at the focal plane of \textcircled{\raisebox{-.9pt} {5}}(\textcircled{\raisebox{-.9pt} {4}}) was recorded by a high-speed camera (Phantom$\textsuperscript{\textregistered}$; \textcircled{\raisebox{-.9pt} {1}}).  We used water and a water/glycerol mixture (glycerol mass fraction 65.2\%) as the liquids. 

\begin{figure}[H]
	\includegraphics[width = 1\textwidth]{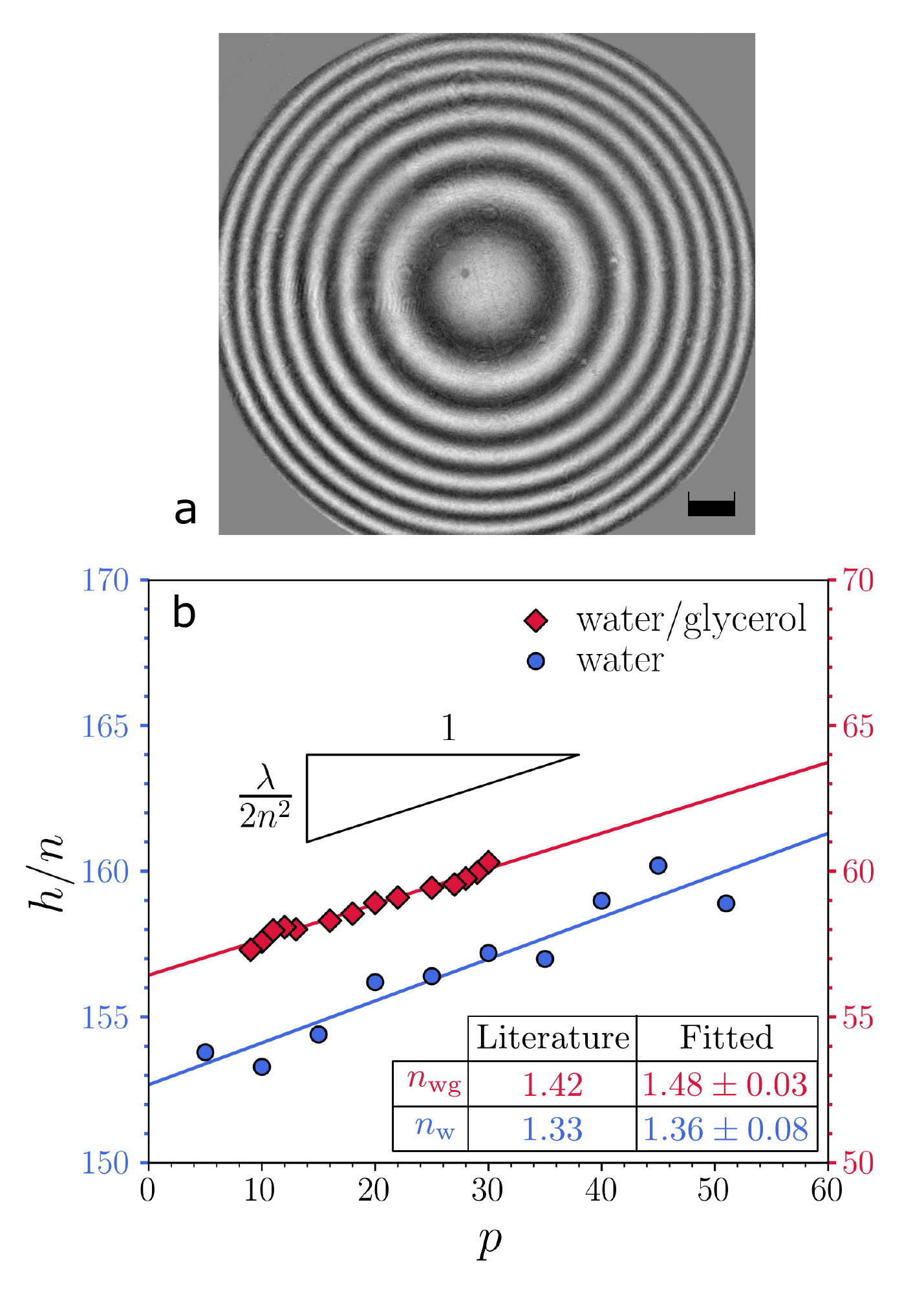}
	\caption{(a) Interference pattern produced by a film of liquid (a water/glycerol mixture: glycerol mass fraction $65.2\%$).  Background subtracted for clarity.  Scale bar: 2mm.  (b) $h/n$ versus accumulated fringe number $p$, for water (blue circles) and a water/glycerol mixture (red squares; glycerol mass fraction $65.2\%$).  Inset: refractive indices deduced from the slopes of the linear fits.}
	\label{fig:hn_p}
\end{figure}

Figure~\ref{fig:hn_p}a shows a typical frame of interference pattern from a high-speed video.  As the thickness of the sample fluctuates, the fringes shrink or expand.  The values of $h/n$ and $p$ (with an arbitrary initial value) have been simultaneously tracked as they vary and are plotted in Fig.~\ref{fig:hn_p}b.  We have performed this for water (shown in blue; 46 fringes appeared from the center in total) and for the water/glycerol mixture (red; 21 fringes appeared from the center in total) separately.  In both cases, the films fluctuated by a few microns in thickness.  The slopes of the linear fits give the refractive indices through Eq.~\ref{eq:hn}.  The measured values compared to those from the literature are shown in the inset of Fig.~\ref{fig:hn_p}.  The results are satisfactory but less precise than those of Table~\ref{tab:h_and_n} due to the additional step of fitting (Eq.~\ref{eq:hp}).

Before ending this subsection, we point out that it is possible to use more than 2 fringes (as in Eqs.~\ref{eq:deconstructive_a0g0}) to remove the dependence on $n$ altogether.  Ishikawa \textit{et al} proposed to use 3 consecutive fringes in a different setup~\cite{ishikawa2004}.  The same idea can potentially be applied here: adding in the fringe of order $m-1/2$ in Eqs.~\ref{eq:deconstructive_a0g0} and cancelling $n$ and $m$ gives:
\begin{align}
	\label{eq:ishikawa}
	\frac{1}{\lambda^2}[\cos2\phi_{m+\frac{1}{2}}+\cos2\phi_{m-\frac{1}{2}}-2\cos\phi_m] = \frac{1}{4h^2}.
\end{align}
In practice, however, we found that this method, along with several other variations derived from the same idea using a triplet of fringes, leads to unacceptably large errors in determining $h$ (that is, using typical values of $n$, $f$ etc. as mentioned above).  This is mainly because the term in the bracket of Eq.~\ref{eq:ishikawa} is extremely small $\sim (\lambda/h)^2$,  hence is extremely sensitive to a measurement error in $\phi_m$.  In comparison, the term in the bracket of Eq.~\ref{eq:hn} is of the order $\sim \lambda/h$, which proves to be large enough to resolve within measurement precision.

\subsection{Parallel sample with tilt ($\alpha=0$, $\gamma \ne 0$)}
\begin{figure}[H]
	\includegraphics[width=0.9\textwidth]{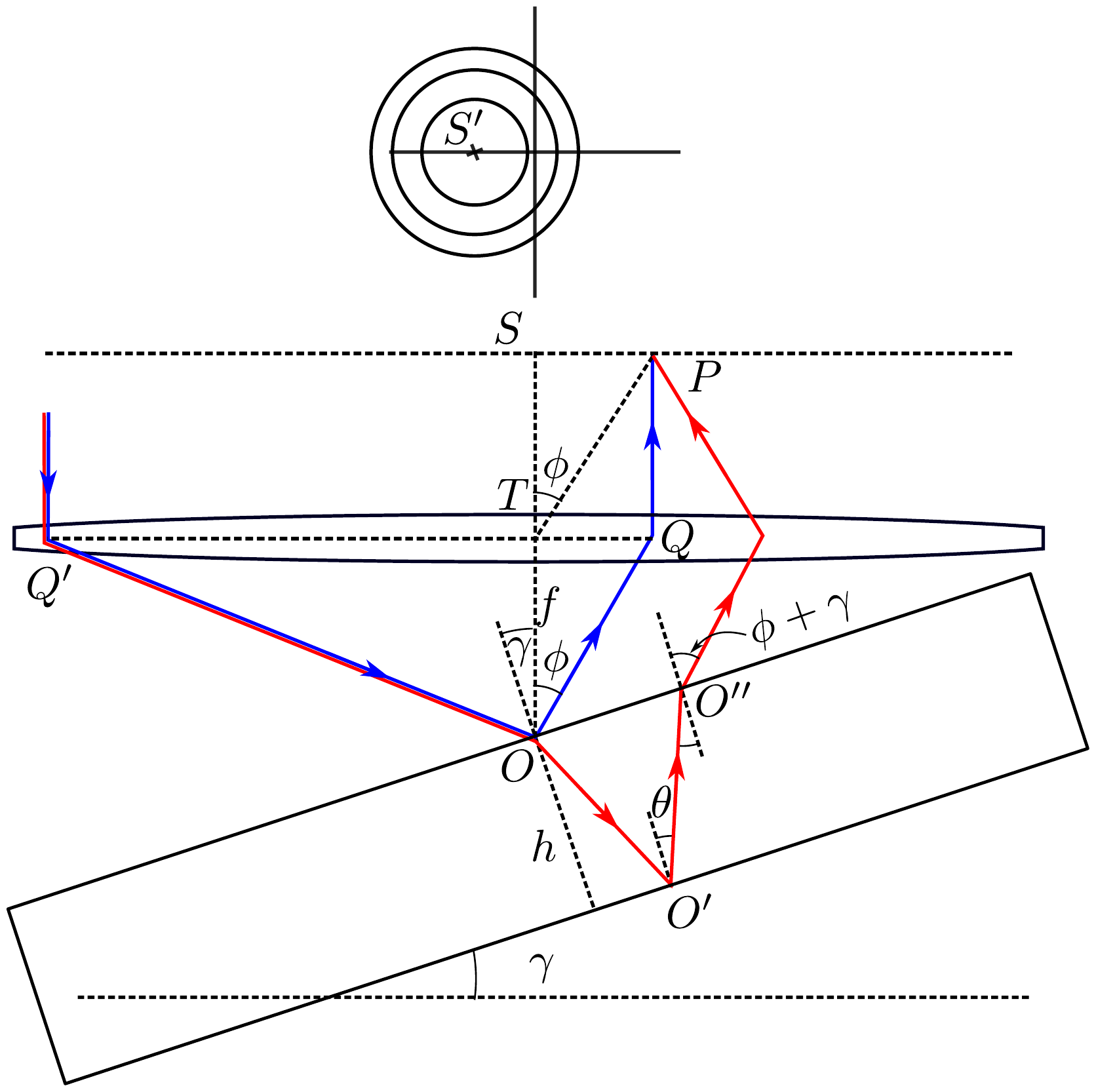}
	\caption{Ray tracing diagram when $\alpha=0,\gamma \ne 0$.  The pattern center $S'$ at the upper focal plane is shifted by $\phi_0=-\gamma$, while the shape and spacing of the fringes remain approximately unchanged.}
	\label{fig:haidinger_a0g1}
\end{figure}

We now consider a sample with parallel sides ($\alpha=0$) and with a slight tilt ($\gamma \ne 0$).  The geometry indicating the plane of the tilt angle $\gamma$ is shown in Fig.~\ref{fig:haidinger_a0g1}.  Due to symmetry the light rays interfering at $P$ all lie strictly in the paper plane.  Snell's law at $O''$ becomes:
\begin{align}
	\label{eq:snell_a0g1}
	\sin(\phi+\gamma) = n\sin\theta.
\end{align}

Since the reflection at $O''$ with $\theta=0$ reaches the pattern center $S'$, the angular position of $S'$, $\phi_0$, can be obtained from substituting $\theta=0$ in Eq.~\ref{eq:snell_a0g1}:
\begin{align}
	\label{eq:phi0}
	\phi_0=-\gamma.
\end{align}
Hence the value of $\gamma$ can be directly obtained by measuring the pattern center shift $\phi_0$.    

Compared with Eq.~\ref{eq:snell_a0g0}, Eq.~\ref{eq:snell_a0g1} indicates that for fringes that are in the paper plane, the fringe position $\phi$ is shifted by the constant $\gamma$.  We next demonstrate that this shift also applies to fringe points out of the paper plane; that is, the pattern shifts as a whole in the focal plane without changing its shape.  

\begin{figure}[htp]
	\includegraphics[width=1\textwidth]{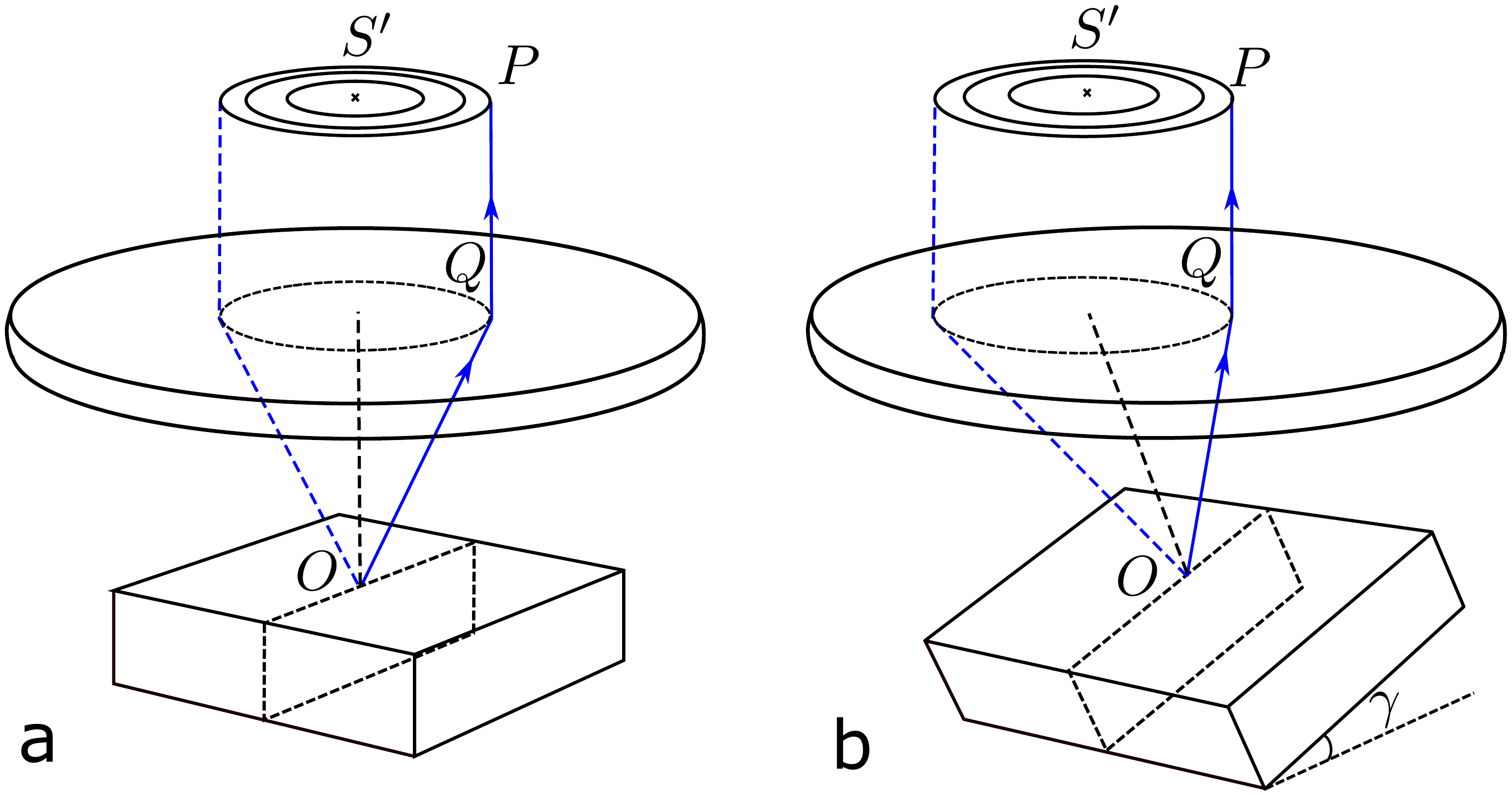}
	\caption{Conical surface formed by the segment of $\overline{OQ}$ when $P$ cycles along a closed fringe. The shape of the cone does not change with the orientation of the sample.  The conic section at $Q$ (shape of the fringe) remains approximately unchanged from a) $\gamma=0$ to b) $\gamma > 0$.}
	\label{fig:cone}
\end{figure}

For all points along the closed contour of an interference fringe, the corresponding set of line segments $\overline{OQ}$ in Fig.~\ref{fig:haidinger_a0g1} form a conical surface with its apex at $O$.  The axis of the cone is perpendicular to the upper surface of the sample at point $O$.  The aperture of the cone is determined by the angle of refraction at $O''$ in Fig.~\ref{fig:haidinger_a0g1} to be $\phi+\gamma = 2\arcsin(n\sin\theta)$, which is independent of the sample orientation $\gamma$.  Consequently, when the slab is tilted by $\gamma$, the cone aperture does not change, while the cone as a whole is tilted by the same amount together with the sample.  This conical shape, and the rotation caused by the sample tilt,  is schematically illustrated in Fig.~\ref{fig:cone}a ($\gamma=0$) and b ($\gamma>0$).

The fringe position is indicated by the intersection of the cone and the lens plane, which is a conic section.  The length of the major axis $2a$ is indicated in Fig.~\ref{fig:haidinger_a0g1} as $\overline{QQ'}$:
\begin{align}
	\label{eq:major}
	2a = \overline{QQ'} &=\overline{QT}+\overline{TQ'}\nonumber\\
	&= f[\tan\phi+\tan(\phi+2\gamma)]\nonumber\\
	&=2f\tan[\arcsin(n\sin\theta)]+O(\gamma^2).
\end{align}
In the last step, Eq.~\ref{eq:snell_a0g1} was used and the result was expanded in power series of $\gamma$.  The length of the minor axis $2b$ can be obtained from the definition~\cite{thomas1974} of the eccentricity $e$ of a conic secion: $e = \sqrt{1-b^2/a^2} = \sin\gamma/\cos(\phi+\gamma)$, from which
\begin{align}
	\label{eq:minor}
	2b &= 2a\sqrt{1-\frac{\sin^2\gamma}{\cos^2(\phi+\gamma)}}\nonumber\\
	&= 2a+O(\gamma^2).
\end{align}
As can be seen from Eq.~\ref{eq:major} and~\ref{eq:minor}, the lengths of the major and minor axes are unaffected by $\gamma$ to first order.  We conclude that the pattern will be merely \textit{translated} in the focal plane by tilting the sample, while the shape and scale of the fringes are unaffected.

As a result, after the transformation $\phi_m \rightarrow \phi_m-\phi_0$, the analysis of the previous section applies here to the case of a tilted sample with parallel sides .  In other words, one can apply the same analysis, taking the pattern center $S'$ as the new origin for angle measurement. (Recall that the screen center $S$ was taken as the origin in the previous section).  We verified in experiment that when the sample was tilted slightly there simply was a translation of the concentric rings; the spacing and shape of the interference pattern was unchanged.

\subsection{Non-parallel surfaces with no tilt ($\alpha\ne 0$, $\gamma=0$)}
\begin{figure}[H]
	\includegraphics[width=1\textwidth]{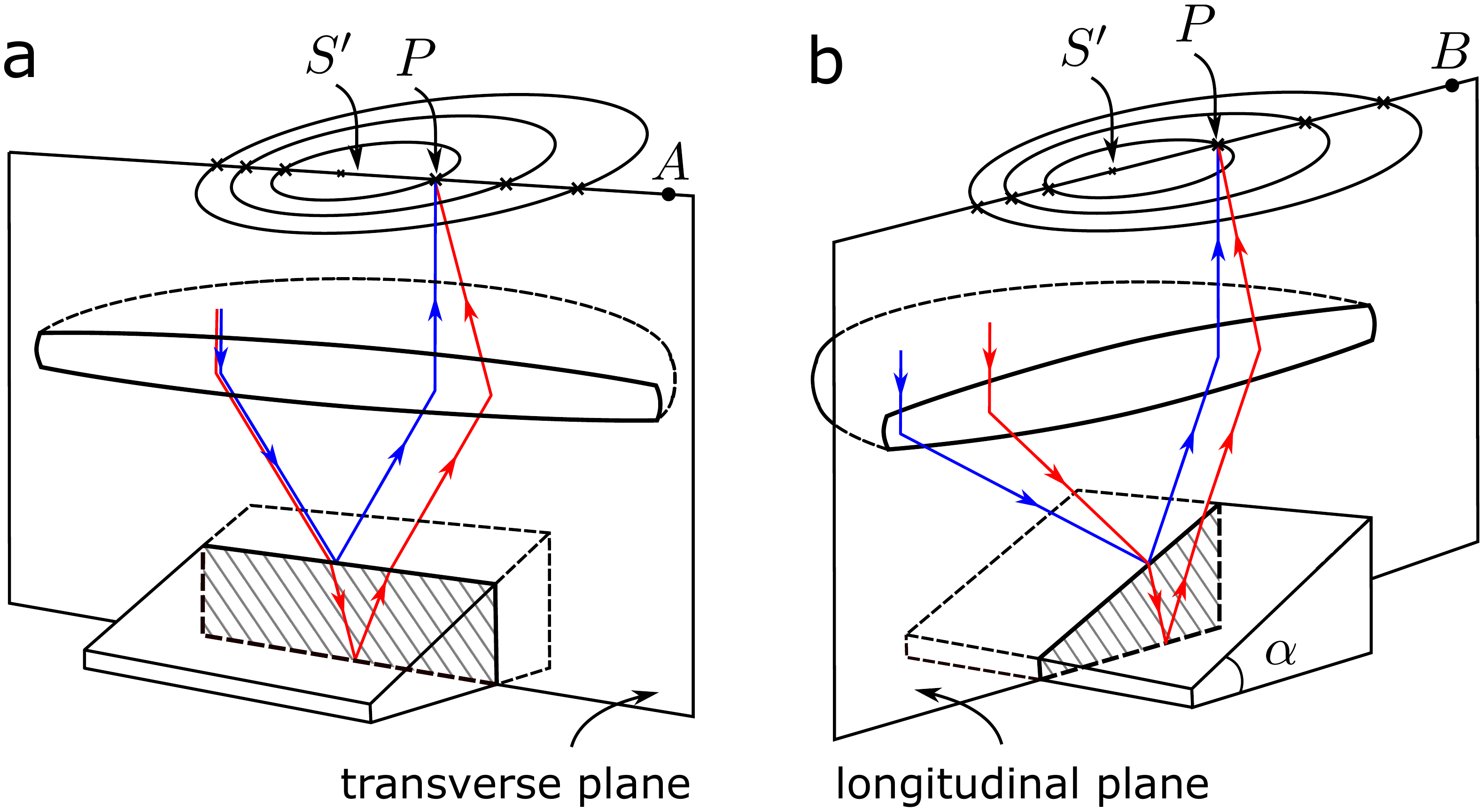}
	\caption{Schematics for $\alpha \ne 0,\gamma=0$.  (a) The transverse plane intersects the sample to form a rectangular cross section (shaded area). Light rays interfering at point $P$ lie approximately in the transverse plane. Analyzing the ray tracing diagram projected onto this plane gives $h$ and $n$.  (b) The longitudinal plane intersects the sample to form a wedge cross section (shaded area) of slope $\alpha$.  Analyzing the ray tracing diagram in this plane gives $h$ and $\alpha$.} 
	\label{fig:crosssection}
\end{figure}

In this section we examine the situation of an untilted ($\gamma=0$) sample in which the two surfaces are not parallel.  Locally it assumes a wedge shape with an opening angle $\alpha\ne0$.  Along the transverse direction  (see Fig.~\ref{fig:crosssection}a) the wedge thickness is a constant while in the longitudinal direction (see Fig.~\ref{fig:crosssection}b) the wedge has the steepest slope, $\alpha$.  We demonstrate below that the wedge thickness, the refractive index and the wedge opening angle can be obtained by studying the spacing of interference fringes along the transverse and longitudinal directions separately.

\subsubsection{transverse direction}
Consider a vertical plane whose intersection with the interference pattern passes $S'$ and is along the transverse direction $\overline{S'A}$, shown in Fig.~\ref{fig:crosssection}a.  The wedge cross section subtended by this plane has a rectangular cross-section as shown by the shaded region in Fig.~\ref{fig:crosssection}a.  We denote this plane as the transverse plane.  Anticipating the result discussed below, the transverse direction of $\overline{S'A}$ can be uniquely identified as the only direction along which the pattern is symmetric about the center $S'$.  In the limiting case of $\alpha=0$, the light rays responsible for the interference at a point $P \in \overline{S'A}$, traced out in Fig.~\ref{fig:crosssection}a, all lie in this transverse plane.  When $\alpha$ is slightly increased to a small non-zero value, the shown light rays deviate from the transverse plane with the same order of smallness as $\alpha$.  We demonstrate in the Appendix that when such \emph{nearly coplanar} light rays are projected onto a transverse plane, their projections obey the two-dimensional Snell's law to first order in the deviations from that plane.  In other words, in the transverse plane the simple analysis of previous sections still applies as an approximation \emph{for the projections}.  In the situation of Fig.~\ref{fig:crosssection}a, the projections of the light rays responsible for interference along $\overline{S'A}$ onto the transverse plane reduces the geometry to that of Fig.~\ref{fig:haidinger_a0g0}.  Applying the method of Section A to the fringe positions along $\overline{S'A}$ gives $h$ and $n$.  

\subsubsection{longitudinal direction}
\begin{figure}[htp]
	\includegraphics[width=0.9\textwidth]{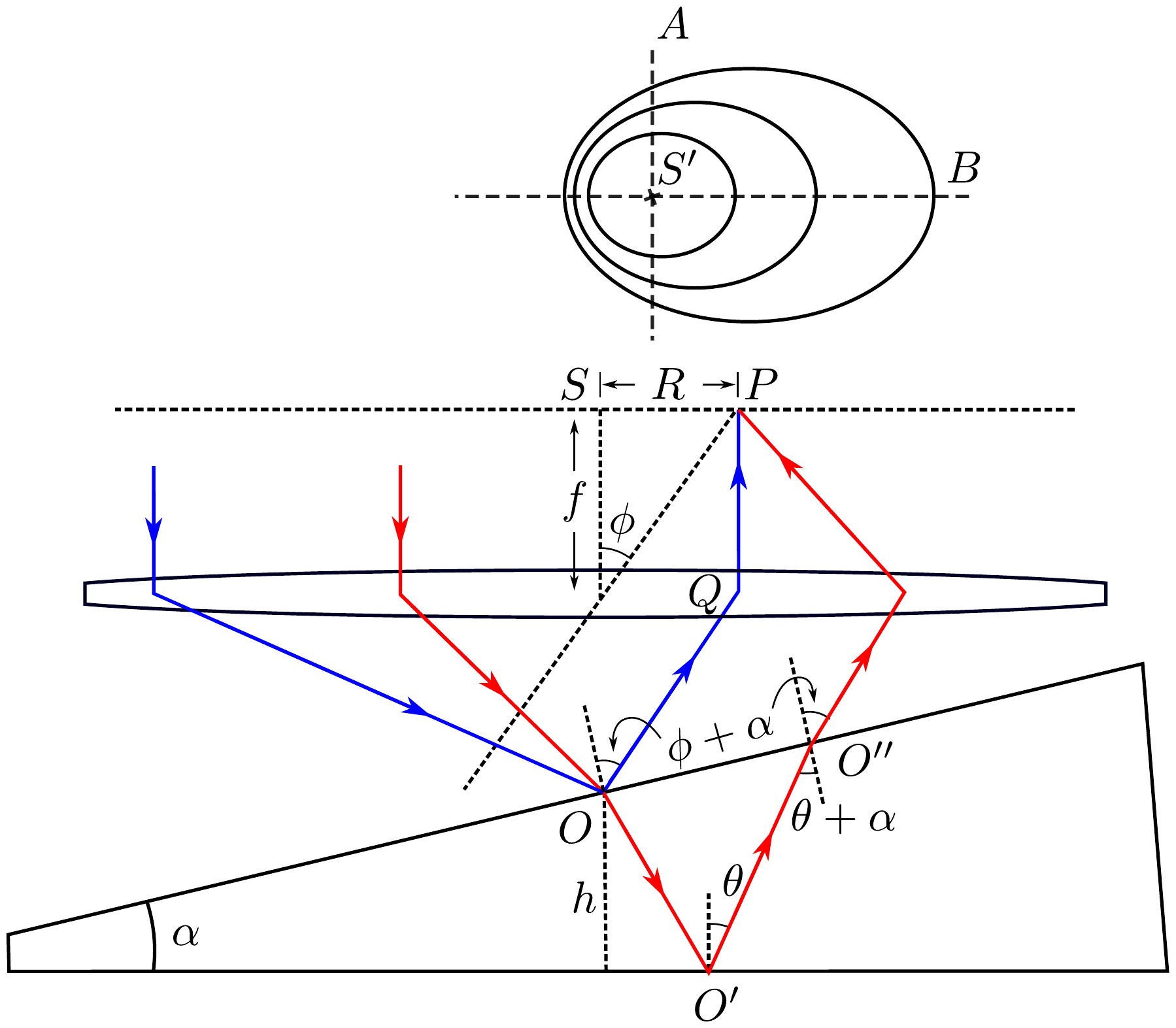}
	\caption{Ray tracing diagram when $\alpha \ne 0,\gamma = 0$.  The pattern center $S'$ at the upper focal plane is shifted by $\phi_{0,\alpha}$ (Eq.~\ref{eq:phi0_a1g0}), while the shape of the fringes are also skewed. $\overline{S'B}$: the longitudinal direction; $\overline{S'A}$: the transverse direction.}
	\label{fig:haidinger_a1g0}
\end{figure}

In the longitudinal plane shown in Fig.~\ref{fig:crosssection}b, the cross section of the wedge makes a trapezoid with opening angle $\alpha$ (shaded region).  The fringe positions in this plane contain the information about the wedge opening angle, $\alpha$.  Note now the traced light rays all lie strictly in this perpendicular plane.  Figure.~\ref{fig:haidinger_a1g0} shows the detailed geometry of this cross section.  

Snell's law at $O''$ becomes:
\begin{align}
	\label{eq:snell_a1g0}
	\sin(\phi+\alpha) = n\sin(\theta+\alpha).
\end{align}
As we can see, the angle $\alpha$ shifts and skews the interference pattern determined by the right-hand-side of Eq.~\ref{eq:snell_a1g0}, if $n \ne 1$.  This is schematically by the shape of the interference patterns shown at the top of Figs.~\ref{fig:crosssection}a, b and Fig.~\ref{fig:haidinger_a1g0}.  The shift of the pattern center, denoted by $\phi_{0,\alpha}$, corresponds to $\theta=0$ in Eq.~\ref{eq:snell_a1g0}: 
\begin{align}
	\phi_{0,\alpha} = \arcsin(n\sin \alpha) -\alpha.
	\label{eq:phi0_a1g0}
\end{align}
We define
\begin{align}
	\label{eq:Phi}
	\Phi_m \equiv \phi_m - \phi_{0,\alpha},
\end{align}
to quantify the angular distance from the stripe to the pattern center.  The motivation for Eqs.~\ref{eq:Phi} is to cancel out the uncertainty of the center position $R=0$ so that our measurements depend on the shape, but not the position, of the interference patterns.

Equations~\ref{eq:destructive},~\ref{eq:constructive} and~\ref{eq:snell_a1g0} lead to a more complicated relation:
\begin{align}
	\frac{\lambda}{4nh} = \cos\bigg(\arcsin\frac{\sin[\Phi_m+\arcsin (n\sin\alpha)]}{n}-\alpha\bigg)\nonumber\\
	-\cos\bigg(\arcsin\frac{\sin[\Phi_{m+\frac{1}{2}}+\arcsin(n\sin\alpha)]}{n}-\alpha\bigg).
	\label{eq:complicated_a1g0}
\end{align}

Equation~\ref{eq:complicated_a1g0} contains the relation between the measurements, $\Phi_m$'s, and the unknowns $h$ and $\alpha$.  However, unlike Eq.~\ref{eq:hn}, these quantities cannot be cleanly separated in Eq.~\ref{eq:complicated_a1g0} to form an explicit mapping.  To proceed, we require that for a set of measurements of $\Phi_m$'s, each adjacent pair should produce through Eq.~\ref{eq:complicated_a1g0} a consistent value of $h$, given a fixed parameter $\alpha$.  More specifically, Eq.~\ref{eq:complicated_a1g0} can be re-written as:  
\begin{align}
	(h_0, h_{\frac{1}{2}}, h_1, \dots)  = f(\Phi_0, \Phi_{\frac{1}{2}}, \Phi_1,  \Phi_{\frac{3}{2}}, \dots; \alpha),
	\label{eq:h_array}
\end{align}
where the thickness $h_0$ is deduced from $\Phi_0$ and $\Phi_{1/2}$, $h_{1/2}$  from $\Phi_{1/2}$ and $\Phi_1$, and so on.  The true value of $\alpha$ should make the left hand side of Eq.~\ref{eq:h_array} a constant series.  We thus seek $\alpha$ that minimizes the standard deviation $\sigma(h_0, h_{\frac{1}{2}}, h_1,\dots)$: 
\begin{align}
	\alpha\textsubscript{fit} = \argmin_{\alpha}\sigma(f(\Phi_0, \Phi_{\frac{1}{2}}, \Phi_1,  \Phi_{\frac{3}{2}}, \dots; \alpha)).
	\label{eq:alpha_fit}
\end{align}
Once $\alpha\textsubscript{fit}$ is obtained, fitted value of thickness $h\textsubscript{fit}$ can be calculated from the mean $\langle h_0, h_{\frac{1}{2}}, h_1,\dots \rangle$: 
\begin{align}
h\textsubscript{fit} = \langle f(\Phi_0, \Phi_{\frac{1}{2}}, \Phi_1,  \Phi_{\frac{3}{2}}, \dots; \alpha\textsubscript{fit})\rangle .
	\label{eq:h_fit}
\end{align}

As a proof of concept for this part, we carry out a simulation applying this method using the setup shown in Fig.~\ref{fig:setup}.  The supposed sample is of thickness $h=95$ $\mu$m, wedge slope $\alpha=5^{\circ}$, refractive index $n=1.4$.  Angles of $\phi$'s are to be obtained from a high-speed camera by measuring $R$ (Eq.~\ref{eq:geometry}).  The measurement errors are simulated to be normally distributed with variance of 1 pixel$\times$1 pixel, corresponding to a resolution of around 5 $\mu$m and is easily attainable by current high-speed cameras.  Values of $\Phi_m$'s are extracted from 10 bright/dark stripes counting from the pattern center using Eqs.~\ref{eq:Phi}.  We use four simultaneous wavelengths of $\lambda =$ 443, 532, 626, 709 nm to increase the data size so as to reduce the standard errors.  This improvement is more and more significant for thinner and thinner samples as fewer and fewer stripes will show up on the screen.  To achieve this one can use the setup similar to that of ~\cite{he_nagel2019} using a white light source.  The returning white-light beam can be split into multiple synchronized cameras using different color filters.  Using multiple colors also helps to identify the center of the pattern (for Eq.~\ref{eq:Phi}).  

\begin{figure}[H]
	\includegraphics[width=1\textwidth]{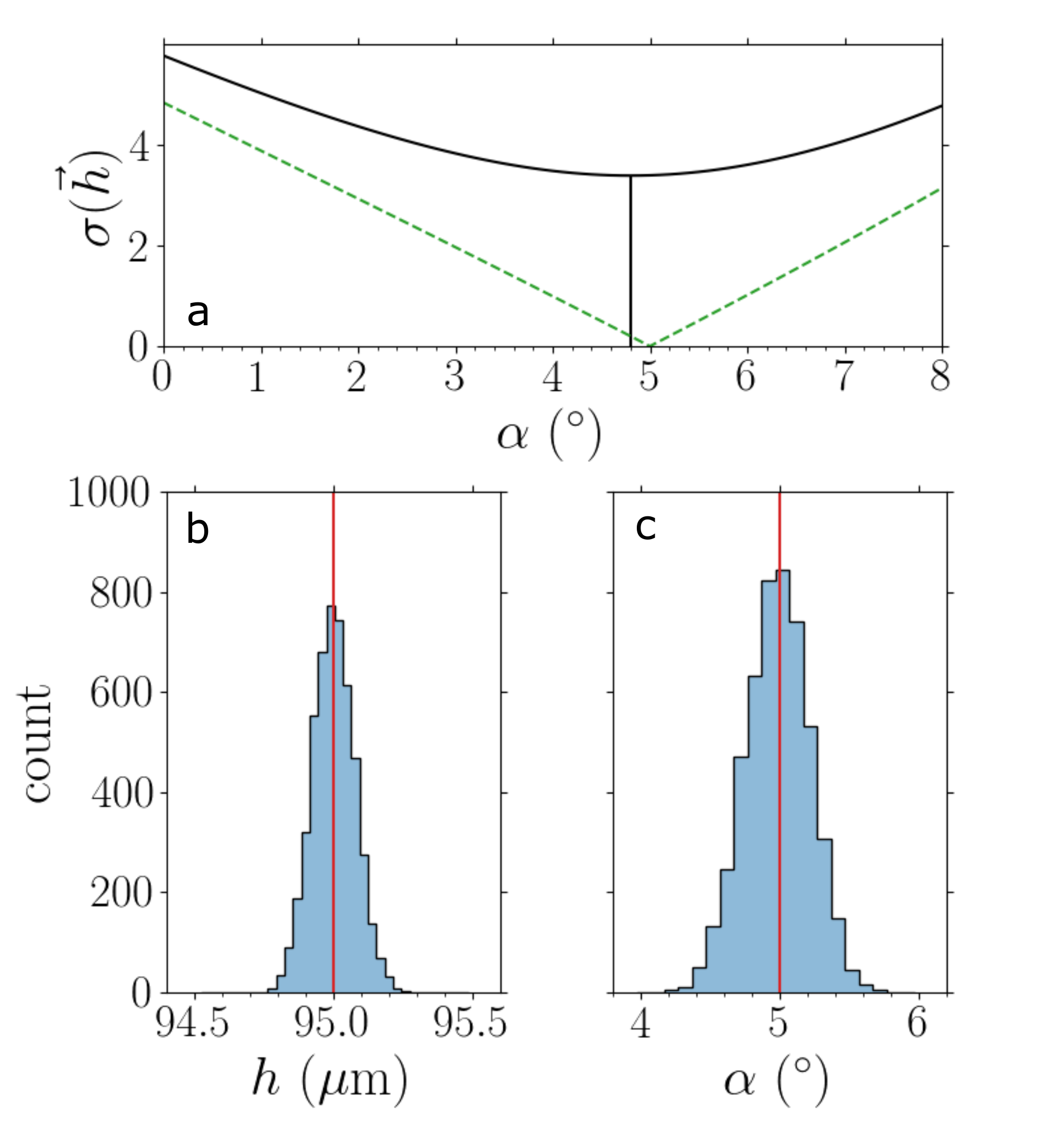}
	\caption{(a) A typical simulated curve of $\sigma(\vec{h})$ as a function of $\alpha$, for a sample of $h=95$ $\mu$m, $\alpha=5^{\circ}$ (solid curve).  The location of the minimum (solid vertical line) indicates the best estimate of $\alpha\textsubscript{fit}=4.8^{\circ}$.  Dashed line: ideal curve without uncertainties.  (b) and (c) Histograms of $h\textsubscript{fit}$ and $\alpha\textsubscript{fit}$, showing mean values (red vertical lines) and widths of the distributions for $N=5000$ simulations.}
	\label{fig:hist}
\end{figure}

The solid curve in Fig.~\ref{fig:hist}a shows the standard deviation $\sigma(h_0, h_{\frac{1}{2}}, h_1,\dots)$ as a function of the parameter $\alpha$, calculated from Eq.~\ref{eq:h_array} for one set of $(\Phi_0, \Phi_{\frac{1}{2}}, \Phi_1,  \Phi_{\frac{3}{2}}, \dots)$ with simulated uncertainties.  The minimum position gives a fitted value of $\alpha\textsubscript{fit}=4.8^{\circ}$, and hence $h=94.97$ $\mu$m (not shown), which is consistent with $\alpha=5^{\circ}$, $h=95$ $\mu$m.  The dashed line shows the same function, but without uncertainties.  We can see that the effect of measurement noise is to reduce the depth and sharpness of the minimum, as well as to induce a small deviation to the location of the minimum.  We note that minimizing $\sigma$ as a function of $\alpha$ does not require any special algorithm: $\alpha$ is a bounded quantity $\in (0^{\circ}, 90^{\circ})$ so a brute force approach is always possible.

To test the robustness of this method, we regenerate the errors for $N=5000$ times and collected the fitted results of $h\textsubscript{fit}$ and $\alpha\textsubscript{fit}$ in Fig.~\ref{fig:hist}b and c.  From this we can conclude that $h\textsubscript{fit}=95.00\pm0.05$ $\mu$m and $\alpha\textsubscript{fit}=5.0\pm0.2^{\circ}$.

In sum, we have first made use of the transverse direction, $\overline{S'A}$, of the interference pattern to obtain $n$ and $h$, and then the longitudinal, $\overline{S'B}$, to obtain $h$ and $\alpha$.  Conveniently, if measuring a known material (with a known $n$), only the fringe positions along longitudinal $\overline{S'B}$ is needed to obtain $h$ and $\alpha$.  In any case, we effectively used the \textit{eccentricity} of the interference pattern to deduce the angle of $\alpha$.    

\subsection{Tilted ($\gamma \ne 0$) Wedge ($\alpha\ne 0$)}
\begin{figure}
	\includegraphics[width=0.9\textwidth]{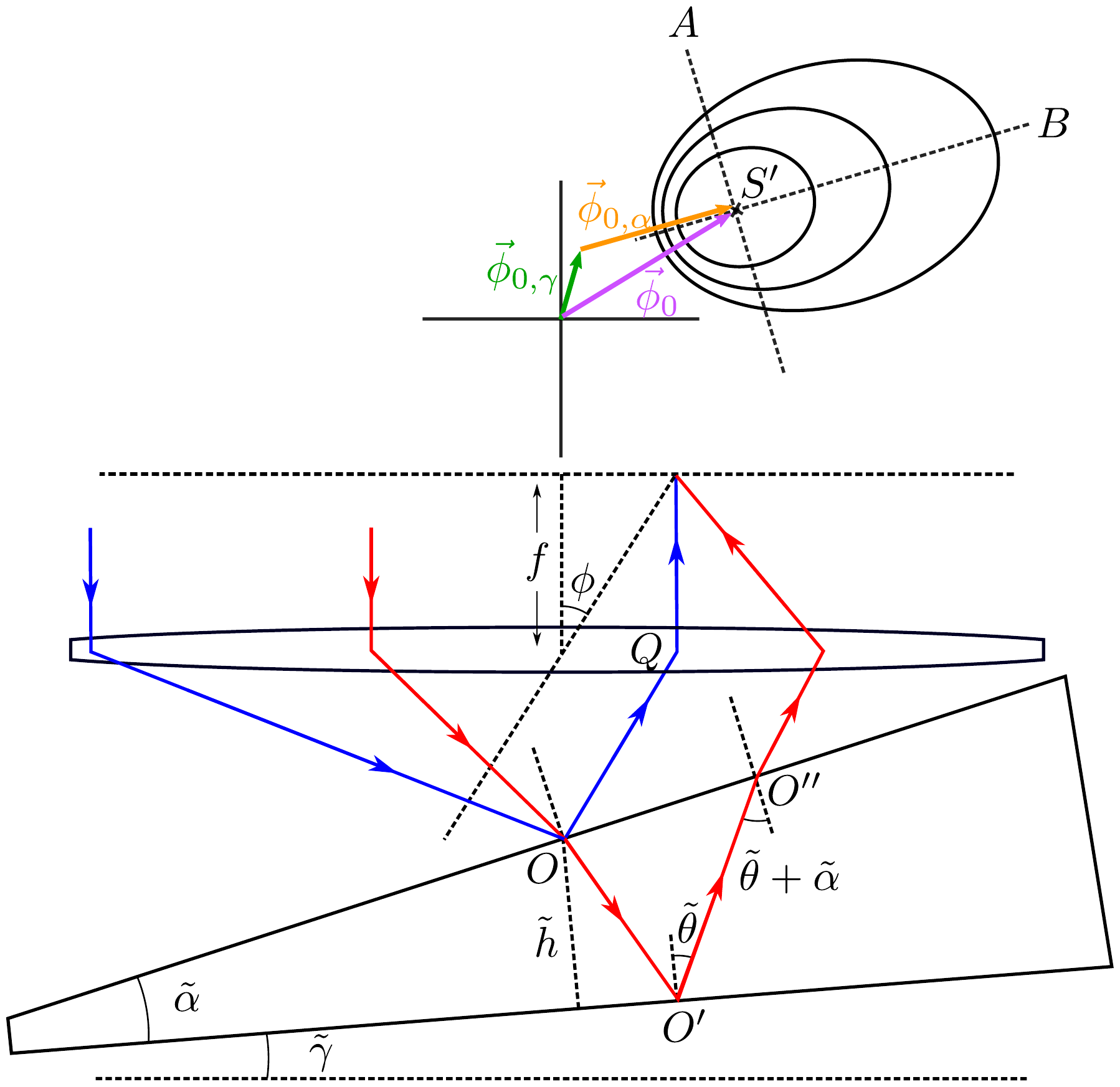}
	\caption{Projected ray tracing diagram when $\alpha \ne 0,\gamma \ne 0$.  Tilded symbols: corresponding projected values.  $\gamma\ne0$ shifts the pattern by $\vec{\phi}_{0,\gamma}$ (green).  $\alpha\ne0$ shifts the pattern by $\vec{\phi}_{0,\alpha}$ (orange) and skews the shape of the fringes.  The net shift of the pattern $\vec{\phi}_0$ (purple) is the vector sum of $\vec{\phi}_{0,\alpha}+\vec{\phi}_{0,\gamma}$.}
	\label{fig:haidinger_a1g1}
\end{figure}

Now we are in the position to deal with the most general case of both $\alpha \ne 0$ and $\gamma \ne 0$, with the directions of the two angles not necessarily in the same plane.  Fig.~\ref{fig:haidinger_a1g1} shows the projection of the ray-tracing diagram onto a plane passing the pattern center $S'$.  Note that now in general the projected values of $\tilde{h}$, $\tilde{\alpha}$, $\tilde{\gamma}$ and $\tilde{\theta}$ differ from the original values of $h$, $\alpha$, $\gamma$ and $\theta$.  However, we will show below that one need not obtain these projected values to deduce the original values.

As with the case in Section C, $\alpha$ shifts and skews the otherwise circular, concentric pattern.  Similar to the discussion in Section B, $\gamma$ linearly translates the whole (whether skewed or not) pattern in the upper focal plane.  The latter can be seen from an examination of the cone surface formed by $\overline{OQ}$ as done in Section B.  The relative orientation and angle of the cone again does not depend on the wedge tilt $\gamma$ (although the cone is now oblique).  Similar to the case of Section B, the intersection of the cone and the lens plane, indicating the fringe shape, is approximately independent of $\gamma$.  The resultant pattern position is a vector composition of the effects of $\alpha$ and $\gamma$.

The use of Eq.~\ref{eq:Phi} becomes crucial, for it makes the analysis independent of the origin position.  Even though the pattern is shifted, one can apply the same method of Section C to obtain the value of $\alpha$.  

After obtaining $\alpha$, the translation due to $\alpha$ follows from Eq.~\ref{eq:snell_a1g0}:
\begin{align}
	\vec{\phi}_{0,\alpha} = \arcsin(n\sin \alpha) -\alpha,
\end{align}
whose direction is along the direction of $\overline{S'B}$ (orange arrow in Fig.~\ref{fig:haidinger_a1g1}).  Suppose the position of $S'$ is $\vec{\phi_0}$, the direction and magnitude of the shift due to $\gamma$ (green arrow in Fig.~\ref{fig:haidinger_a1g1}) can be found from:
\begin{align}
	\vec{\phi}_{0,\gamma}=\vec{\phi_0}-\vec{\phi}_{0,\alpha}, 
\end{align}
which is of the same size but the opposite direction of $\gamma$ (Eq.~\ref{eq:phi0}).

\section{Conclusions}
In this work we presented a method to measure the height, tilt, opening angle and index of refraction of a thin wedge using interference techniques.  By using a high-speed camera to capture the fluctuating interference fringes, the refractive index of the material can be accurately estimated.  Once the refractive index is known, the absolute thickness, the wedge slope and the local thickness gradient can be identified from a \textit{single} frame.  This method requires simple instrumentation and gives robust results.  While this is an improvement over and extension to solid thin film measurements, it is particularly useful in dynamic experiments of soft materials where the thickness and tilt of the film vary rapidly over time.


Our method is robust. We found in experiments that if the liquid interface is slightly displaced from the focal point of the convex lens, the measurement results remain virtually unchanged.  This is important since it is impossible for the interface to always remain at the focal point if it is also fluctuating in time.  If one is only interested in measuring the film thickness, ignoring the effect of $\alpha$ altogether would not change the result of $h\textsubscript{fit}$ much.  This latter is reflected in Fig.~\ref{fig:hist}b and c:  when $\alpha\textsubscript{fit}$ changes by $\sim 16\%$, $h\textsubscript{fit}$ only changes by $\sim 0.2\%$.  Moreover, this method only relies on a geometrical measurement, the shape of the fringes, instead of their intensity.  This offers a different method, and makes this method potentially more robust, compared with optical intensity techniques such as ellipsometry~\cite{tompkins2005handbook, fujiwara2007spectroscopic} and photo-absorption (\textit{e.g.}~\cite{bischofberger2014,videbaek2019}).  It simplifies the setup, the analysis,  as well as the experimental procedure as no calibration of intensity is required beforehand.  Despite its non-essential role, the pattern intensity can still be used to detect the shape of the fringes by pattern fitting algorithms.

We note that only in Section C.~2 the analysis is exact.  Various approximations have been used in Section A (Eqs.~\ref{eq:deconstructive_a0g0_2}; small angle approximation), Section B (Eqs.~\ref{eq:major} and~\ref{eq:minor}), Section C.~1 (``coplanarity approximation'') and related parts in Section D.  The resulting $h\textsubscript{fit}$, $n\textsubscript{fit}$, \dots, are only correct to the first order of the small angles.  


\begin{figure}[htp]
	\includegraphics[width=0.5\textwidth]{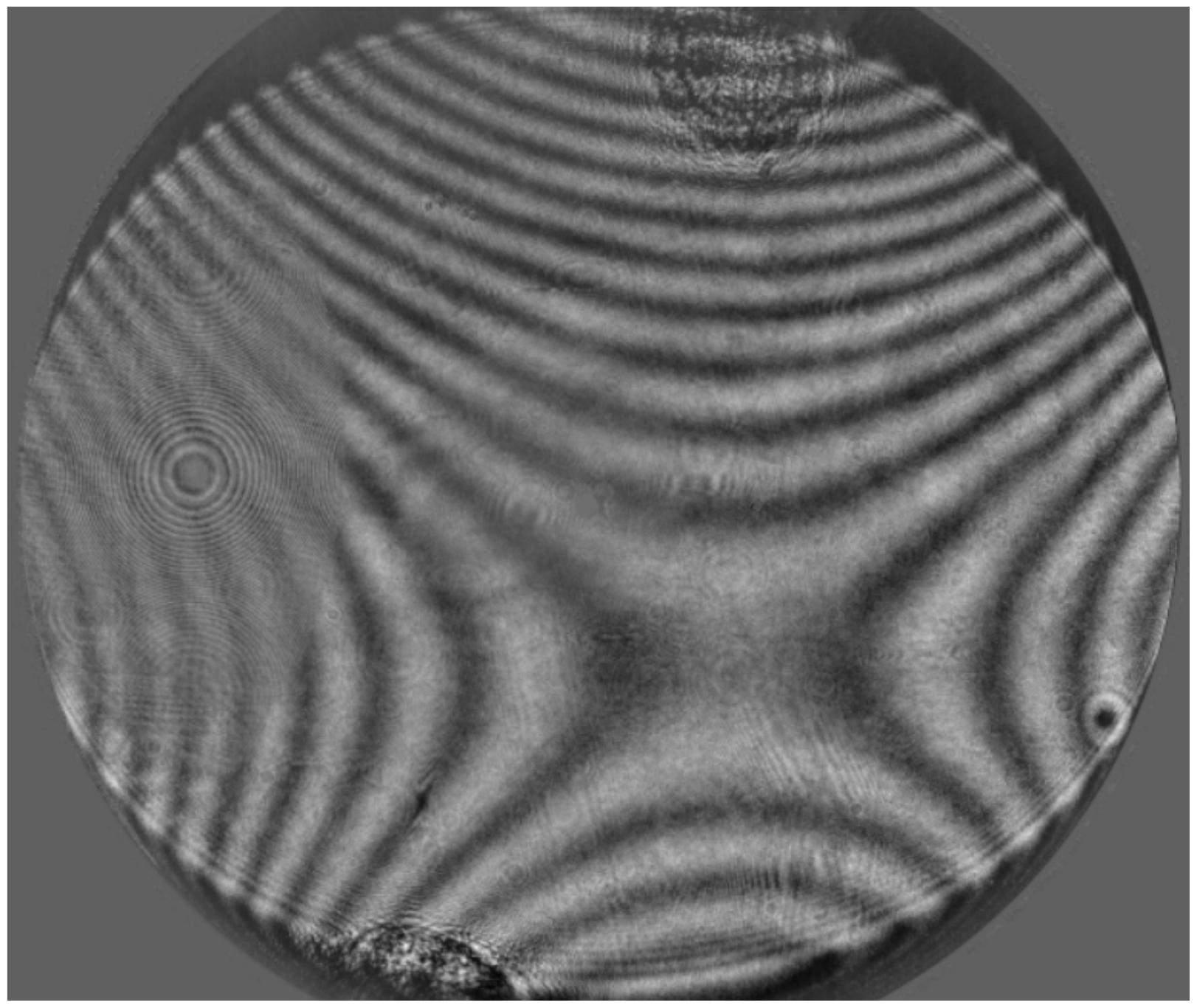}
	\caption{An observed fringe configuration, due to curvature of the sample surface, that cannot be modeled by the simple analysis of this work.  Background subtracted for clarity.}
	\label{fig:saddle}
\end{figure}

This work did not take into account the curvature of the sample interface.  An example of an interference pattern that could not be analyzed by this technique is shown in Fig.~\ref{fig:saddle}, where the fringes are no longer simple closed loops.  A synchronized camera focusing on the liquid film to track the location of the measurement shows that this often occurs close to the edge of the liquid layer or in very curved regions.  
We note that extending the analysis to apply to curved interfaces would be valuable and have wide applications.

\section{Appendix}
In this Appendix we demonstrate that the form of Snell's law is approximately valid for the projection of the ray-tracing diagram onto a plane that is slightly rotated away from the incidence plane. 

\begin{figure}[htp]
	\includegraphics[width=0.3\textwidth]{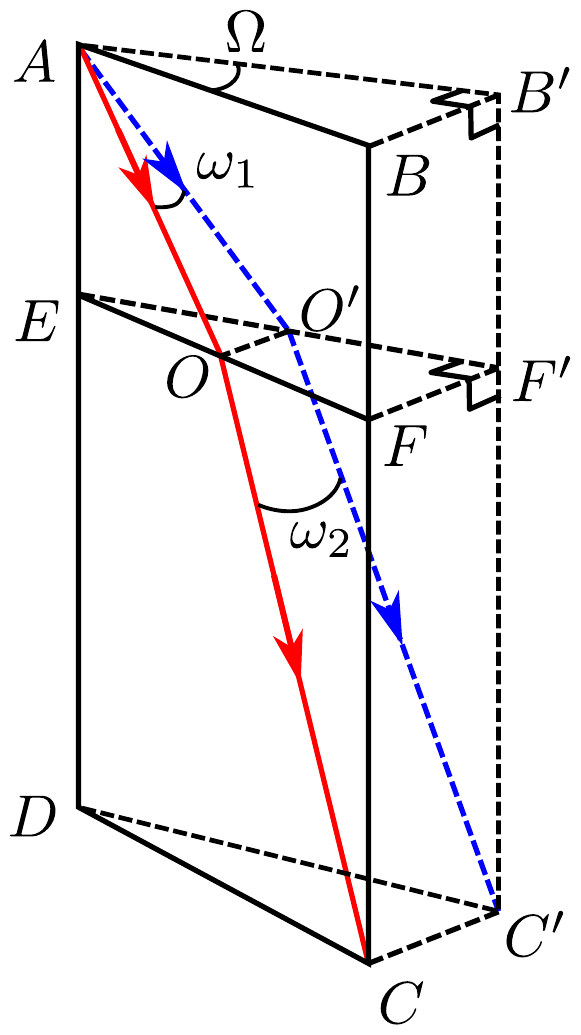}
	\caption{A ray tracing diagram in plane $ABCD$ (red solid arrows) is projected onto a nearby plane $AB'C'D$ (blue dashed arrows).  The projection obeys Snell's law to the first order of the angle $\Omega$ between the planes.}
	\label{fig:snell}
\end{figure}

Consider a plane of incidence $ABCD$, as shown in Fig.~\ref{fig:snell}, where the $\overline{AO}$ and $\overline{OC}$ (red solid lines) indicate a light beam refracted by the interface $\overline{EF}$.  If we set $\overline{OE}=\overline{OF}$, Snell's law is equivalent to: 
\begin{align}
	\frac{\overline{AO}}{\overline{OC}} = \frac{n_1}{n_2},
	\label{eq:incidence_plane}
\end{align}
where $n_1$ and $n_2$ are the refractive index above and below interface $\overline{EF}$.

Onto a plane $AB'CD'$ (dashed outline) rotated by an angle of $\Omega$ about $\overline{AD}$, the ray tracing is projected to be $\overline{AO'}$ and $\overline{O'C'}$ (dashed blue lines), with the projection of the interface $\overline{EF'}$.  Let $\omega_1$ and $\omega_2$ denote the angles between the original and the projected light rays.  Since $\overline{OO'}$, $\overline{FF'} \perp AB'C'D$, we have $\overline{AO'} = \overline{AO}\cos\omega_1$ and $\overline{O'C'} = \overline{OC}\cos\omega_2$.  Substitute into Eq.~\ref{eq:incidence_plane} we get: 
\begin{align}
	\frac{\overline{AO'}}{\overline{O'C'}} = \frac{n_1}{n_2}\frac{\cos\omega_1}{\cos\omega_2},
	\label{eq:projection_plane}
\end{align}
Without loss of generality we assume $n_1 < n_2$.  It follows that  $\omega_2 < \omega_1 < \Omega$.  Thus, 
\begin{align}
	1-\frac{\Omega^2}{2}+\ldots = \cos\Omega<\cos\omega_1<\frac{\cos\omega_1}{\cos\omega_2}<1.
\end{align}
That is to say, $\cos\omega_1/\cos\omega_2$ is different from $1$ by at most a second order correction in the rotation $\Omega$.  Therefore, to the first order of $\Omega$, Eq.~\ref{eq:projection_plane} takes the form:
\begin{align}
	\frac{\overline{AO'}}{\overline{O'C'}} = \frac{n_1}{n_2},
	\label{eq:projection_plane_firstorder}
\end{align}
which is identical to the form of Snell's law Eq.~\ref{eq:incidence_plane}.

Likewise, when the projection plane is rotated about $\overline{AB}$ by a small angle, the same argument applies.  Any direction of the projection plane can be reached by two steps of rotations about $\overline{AD}$ and $\overline{AB}$ (two of the Euler angles), neither of which will produce an error larger than a second-order correction.  We conclude that the projected ray tracing still obeys Snell's law up to first order in the angle between the projection plane and the plane of incidence.   

\section*{Funding}
This work is primarily supported by National Science Foundation (NSF) MRSEC program (DMR-1420709) and National Institute of Standards and Technology, Center for Hierarchical Materials Design (CHiMaD) (70NANB14H012).

\bibliographystyle{apsrev4-2}
\bibliography{refs}

\end{document}